\documentclass[preprint]{revtex4-1}

\usepackage{graphicx}
\makeatletter
\def\@seccntformat#1{%
  \expandafter\ifx\csname c@#1\endcsname\c@section\else
  \csname the#1\endcsname\quad
  \fi}
\makeatother

\begin{document}

\title{Enormous sample scale-up from nL to $\mu$L \\in high field liquid state dynamic nuclear polarization}

\author{Dongyoung Yoon}
\email[E-mail me at ]{dongyoung.yoon@epfl.ch}
\affiliation{Institute of Physics, $\acute{\text{E}}$cole Polytechnique F$\acute{\text{e}}$d$\acute{\text{e}}$rale de Lausanne, CH-1015 Lausanne, Switzerland}

\author{Alexandros I. Dimitriadis}
\affiliation{Institute of Physics, $\acute{\text{E}}$cole Polytechnique F$\acute{\text{e}}$d$\acute{\text{e}}$rale de Lausanne, CH-1015 Lausanne, Switzerland}
\affiliation{SWISSto12 SA, 1015, Lausanne, Switzerland}

\author{Murari Soundararajan}
\affiliation{Institute of Physics, $\acute{\text{E}}$cole Polytechnique F$\acute{\text{e}}$d$\acute{\text{e}}$rale de Lausanne, CH-1015 Lausanne, Switzerland}

\author{Christian Caspers}
\affiliation{Institute of Physics, $\acute{\text{E}}$cole Polytechnique F$\acute{\text{e}}$d$\acute{\text{e}}$rale de Lausanne, CH-1015 Lausanne, Switzerland}

\author{J$\acute{\text{e}}$r$\acute{\text{e}}$my Genoud}
\affiliation{Institute of Physics, $\acute{\text{E}}$cole Polytechnique F$\acute{\text{e}}$d$\acute{\text{e}}$rale de Lausanne, CH-1015 Lausanne, Switzerland}

\affiliation{Swiss Plasma Center, $\acute{\text{E}}$cole Polytechnique F$\acute{\text{e}}$d$\acute{\text{e}}$rale de Lausanne, CH-1015 Lausanne, Switzerland}

\author{Stefano Alberti}
\affiliation{Institute of Physics, $\acute{\text{E}}$cole Polytechnique F$\acute{\text{e}}$d$\acute{\text{e}}$rale de Lausanne, CH-1015 Lausanne, Switzerland}

\affiliation{Swiss Plasma Center, $\acute{\text{E}}$cole Polytechnique F$\acute{\text{e}}$d$\acute{\text{e}}$rale de Lausanne, CH-1015 Lausanne, Switzerland}

\author{Emile de Rijk}
\affiliation{Institute of Physics, $\acute{\text{E}}$cole Polytechnique F$\acute{\text{e}}$d$\acute{\text{e}}$rale de Lausanne, CH-1015 Lausanne, Switzerland}
\affiliation{SWISSto12 SA, 1015, Lausanne, Switzerland}

\author{Jean-Philippe Ansermet}
\affiliation{Institute of Physics, $\acute{\text{E}}$cole Polytechnique F$\acute{\text{e}}$d$\acute{\text{e}}$rale de Lausanne, CH-1015 Lausanne, Switzerland}

\begin{abstract}

Dynamic nuclear polarization (DNP) enhances nuclear magnetic resonance (NMR) signals by transferring electron spin polarization to nuclei. As DNP requires microwave magnetic fields $\mathrm{B_1}$ strong enough to saturate electron spins, microwave resonators are generally used to achieve a sufficient $\mathrm{B_1}$, at the expense of restricting the sample size. Higher fields improve NMR sensitivity and resolution. However, resonators at 9 T for example can only hold nano-liters (nL). Larger volumes are possible by avoiding resonators, but the higher power needed to reach $\mathrm{B_1}$ is likely to evaporate the sample. Here, we demonstrate a breakthrough in liquid state DNP at 9 T, boosting the sample size to the $\mu$L range. We could use high-power (70 W) microwaves thanks to a planar probe designed to alleviate dielectric heating. We enhanced the ${}^1$H NMR signal intensity of 2 $\mu$L of liquid water by a factor of 14, while maintaining the water temperature below 40 ${}^{\circ}$C.

\end{abstract}

\maketitle

Nuclear magnetic resonance (NMR) is a powerful analytical tool in chemistry and biochemistry. However, the low sensitivity of NMR, which is directly linked to the strength of the nuclear magnetic moments, generally requires samples containing a large number of nuclear spins, more than $10^{17}$ spins, or long measurement times. Dynamic nuclear polarization (DNP) can increase NMR signal intensity by transferring the much higher polarization of unpaired electron spins to bulk nuclei\cite{Abragam}. DNP requires microwaves (MW) of sufficiently strong magnetic fields ($\mathrm{B_1}$) to saturate the electrons spin resonance. At X band ($\simeq$ 9 GHz at $\simeq$ 3 kG), this can be easily achieved by using microwave resonators with a high quality factor (Q)\cite{Armstrong2009,Hofer2008,Tuerke2012}. The sample is located at a position of low electric field ($\mathrm{E_1}$), thus preventing dielectric heating of the sample under test. However, the size of the microwave resonators decreases as the magnetic field and the  frequency increase. As a consequence, the sample volume contained at the $\mathrm{B_1}$ maximum in microwave resonators is inherently limited by the microwave wavelength. This can restrict the sample volume for DNP at 9 T to nanoliters only. Furthermore, it becomes difficult to combine it with the resonator used for NMR at radio frequencies (RF). 
\begin{figure*}[!htb]
\hspace{-0.07cm}
\centerline{\includegraphics[width=17cm,height = 7cm]{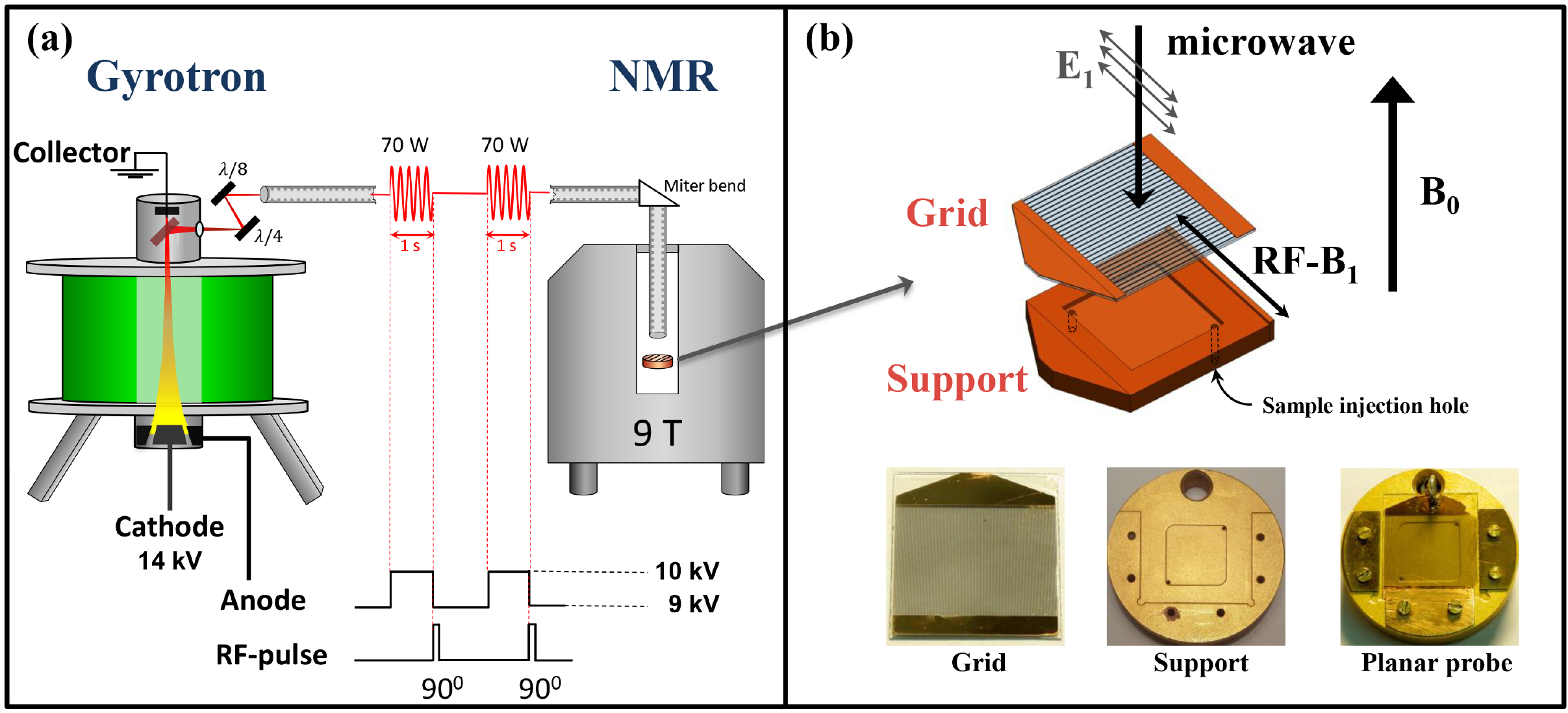}}
\caption{(a) DNP spectrometer composed of a triode-gun gyrotron, $\lambda/4$ \& $\lambda/8$ polarizer mirrors, a 5 m-long corrugated waveguide, a miter bend, a superconducting magnet (9 T), and a planar probe. The cathode voltage is fixed, while the anode voltage is modulated by an external trigger. FID pulse for NMR is turned on immediately after switching off the microwaves. The given cathode, anode, RF-pulse, and gyrotron output correspond to the experimental conditions used for ${}^1$H DNP with water. (b) Planar probe. $\mathrm{B_0}$, $\mathrm{E_1}$, and RF-$\mathrm{B_1}$ represent external static magnetic field, electric field of microwaves, and RF-magnetic field, respectively.}\label{Experiment-set-up}
\vspace{-0.02\textwidth}
\end{figure*} Therefore, microwave resonators are generally not used for high-field DNP ($>$ 5 T) and instead, high-power microwave generators such as gyrotrons or extend interaction klystron (EIK) are used\cite{Becerra1993,Rosay2010,Kemp2016}. A power level of a few watts is sufficient in solid state DNP with magic angle spinning (MAS) at 9 T \cite{Bruker} to saturate at least partially the electron spin resonance in glassy frozen solutions because, in these samples, the electron spin-lattice relaxation time $T_{1e}$ is quite long, about 0.1-1 ms\cite{Thurber2010}. In addition, dielectric heating is weak in solid state DNP because of the relatively low loss tangent (tan $\mathrm{\delta}$ $\simeq$ 0.01) of frozen samples\cite{Nanni2011}. Furthermore, the samples are thermalized thanks to the high cooling power of the cold $\mathrm{N_2}$ used to spin the rotors. Signal enhancement in solid state DNP with MAS can be optimized by choosing suitable radicals\cite{Hu2004,Zagdoun2013}.

\begin{figure*}[!htb]
\centerline{\includegraphics[width=17cm,height = 7cm]{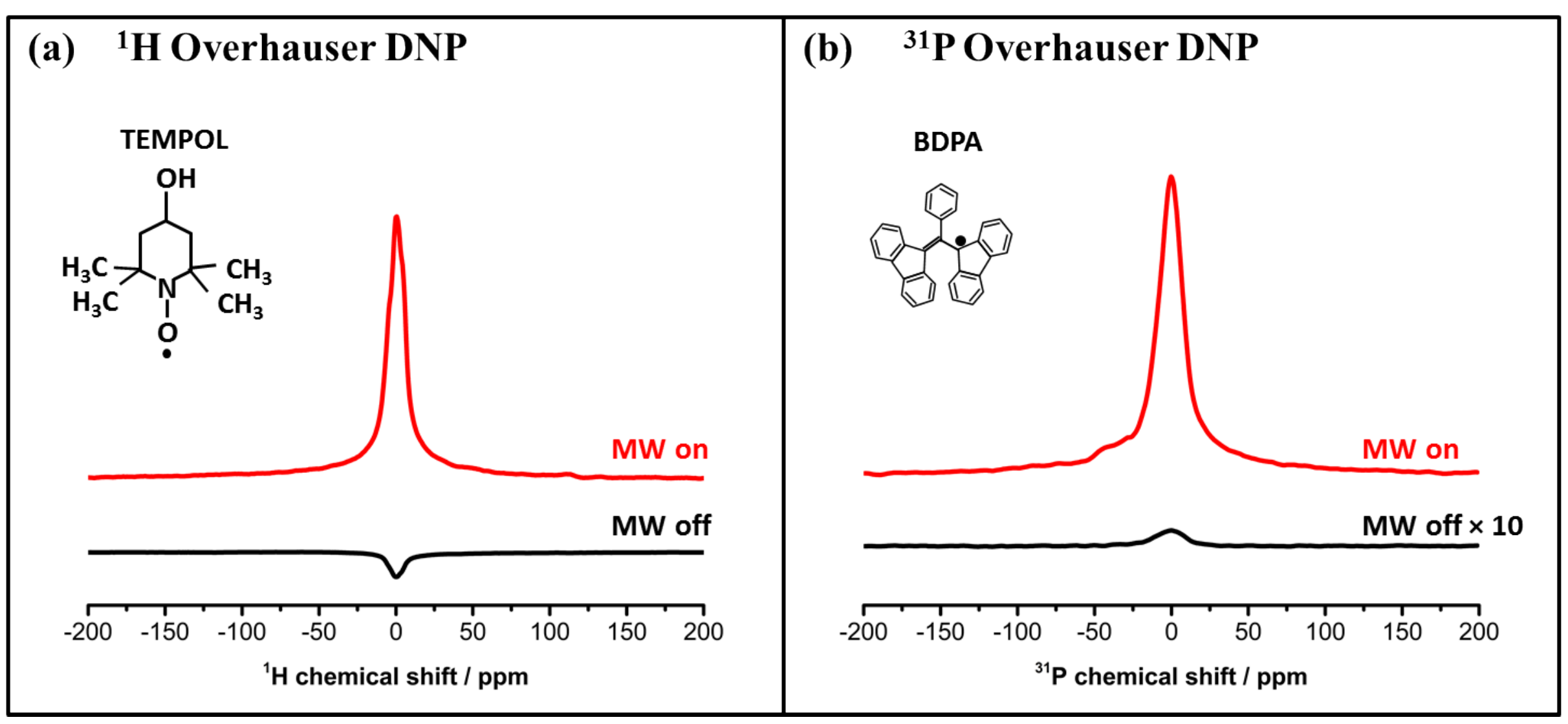}}
\caption{(a) Liquid state ${}^1$H spectra obtained by the planar probe with and without microwaves with 80 mM TEMPOL dissolved in 2 $\mu$L of water. The DNP spectrum was obtained with 70 W of microwave power, 1 s of irradiation time, and 5 s of cooling time (4 cycles). (b) Liquid state ${}^{31}$P DNP spectra obtained by the planar probe with and without microwaves with BDPA dissolved in a solution of 10 $\mu$L fluorobenzene with P$\mathrm{h_3}$P. 17 W of microwave power, 5 s of irradiation time, and 15 s of cooling time were applied for the DNP spectrum (4 cycles).}\label{DNPspectrum}
\vspace{0\textwidth}
\end{figure*}

Saturating electron spins of radicals in liquids at high field is more challenging because $T_{1e}$ is much shorter (\textit{e.g.} $\sim$ 120 ns for TEMPOL radicals)\cite{Prandolini2009}. Therefore, a much larger $\mathrm{B_1}$ is required in liquids than in solids. Since aqueous or other polar solutions generally have larger dielectric loss than frozen solutions (tan $\mathrm{\delta}$ $\geq$ 0.8), more cooling power is needed. Prisner \textit{et al.} conducted liquid state DNP at 9 T using an $e-{}^1$H double resonance probe with a Q $\simeq$ 400 for \textit{e}, and the sample was restricted to only a few nL \cite{Prandolini2008, Prandolini2009,Gafurov2012,Neugebauer2013}. The microwave resonator allowed them to achieve 90 \% of the full electron spin saturation with 100 mW of microwave power. However, dielectric heating large enough to heat the sample over the boiling point could occur, so that active cooling obtained by blowing $\mathrm{N_2}$ gas on the sample was necessary to keep the sample temperature under control.

\begin{figure*}[!htb]
\centerline{\includegraphics[width=16cm,height = 8.5cm]{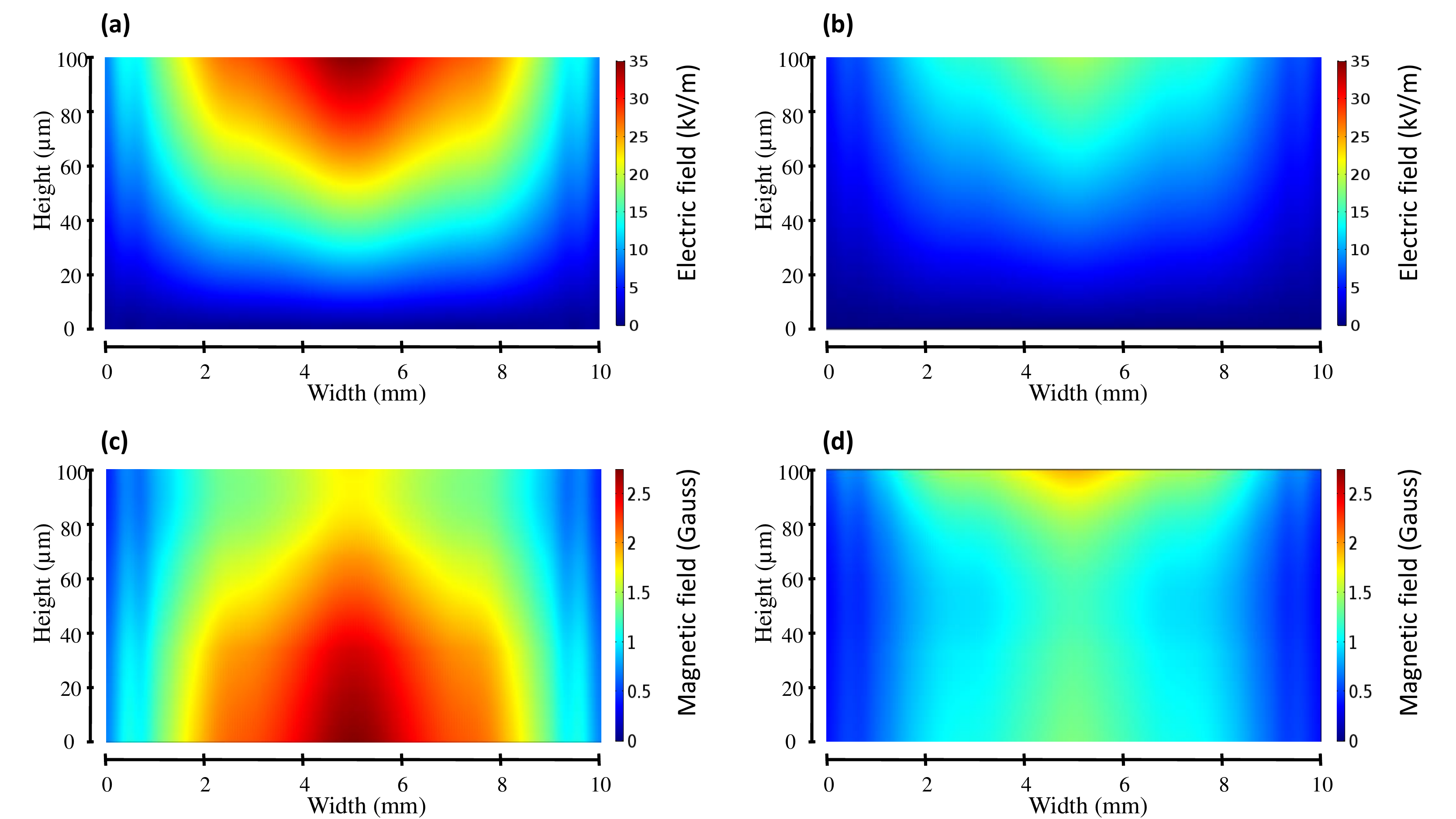}}
\caption{Simulation results on the two dimensional MW electric and magnetic fields (260.5 GHz) in the sample region of the planar probe composed of the grid and the Cu support. The top line of the figures represents the bottom of the fused silica substrate, and the bottom line represents the ground plane of the Cu support. The results are based on the linearly polarized microwaves of 70 W whose electric field is perpendicular to the wires of the grid. (a) MW electric field distribution with $\epsilon^{'}$ = 0.6 and tan $\mathrm{\delta}$ = 0.8. (b) MW electric field distribution with $\epsilon^{'}$ = 0.5 and tan $\mathrm{\delta}$ = 1.2. (c) MW magnetic field distribution with $\epsilon^{'}$ = 0.6 and tan $\mathrm{\delta}$ = 0.8. (d) MW magnetic field distribution with $\epsilon^{'}$ = 0.5 and tan $\mathrm{\delta}$ = 1.2.}\label{simulation}

\vspace{0\textwidth}
\end{figure*}

In this study, we report a new approach that can boost the possible sample volume to microliters for liquid state DNP at 9 T. We used a high-power gyrotron ($\simeq$ 150 W), and obtained DNP without relying on a microwave resonator to saturate electron spins. Such a high MW power would boil off solutions held in conventional NMR probes. An innovative DNP-NMR probe design (Pat. Pend. \cite{annino2013}) prevents heating of the sample because the liquid is located where $\mathrm{E_1}$ is minimum and because the liquid is also well thermalized with the metal backing of the probe. The thin sample (d = 100 $\mu$m $\ll$ $\lambda$/4) can be of as much as 10 $\mu$L. We succeeded in an enhancement of ${}^1$H NMR by a factor of -14 for an aqueous solution and an enhancement of ${}^{31}$P NMR by a factor of 200 for a solution of triphenylphosphine in fluorobenzene. All DNP experiments here were performed without active cooling devices.

\section{Results}

\textbf{DNP spectrometer and experiment procedure.}~~Figure \ref{Experiment-set-up} (a) shows the liquid state DNP spectrometer used in this study, which is composed of a high-power gyrotron, two corrugated polarizing $\lambda$/4 \& $\lambda$/8 mirrors that allow us to control the polarization of the millimeter wave, a 5 m-long corrugated waveguide, a miter bend, a superconducting NMR magnet (9 T), and the planar probe. The gyrotron was designed using a triode-magnetron-injection gun that enables an independent control of the anode voltage\cite{Alberti2012, Alberti2013}. This allows for unique features such as fast frequency tunability and fast switchability\cite{Yoon2016}. The gyrotron is tunable within about 1 GHz around 260 GHz by changing the magnetic field. The gyrotron power shows a maximum of about 150 W near the lowest frequency of its tuning range, and decreases with increasing frequency. Modulation of the anode voltage by an external trigger enables the MW to be switched on and off rapidly, as depicted in Fig. \ref{Experiment-set-up} (a). NMR signals with MW were obtained  as follows: first, MW is turned on for 1-5 s, then a free induction decay (FID) is acquired after MW is turned off. MW is kept off for a while after the NMR signal acquisition to allow the sample to return to the initial temperature.

\begin{figure}[htb]
\centerline{\includegraphics[width=0.55\textwidth]{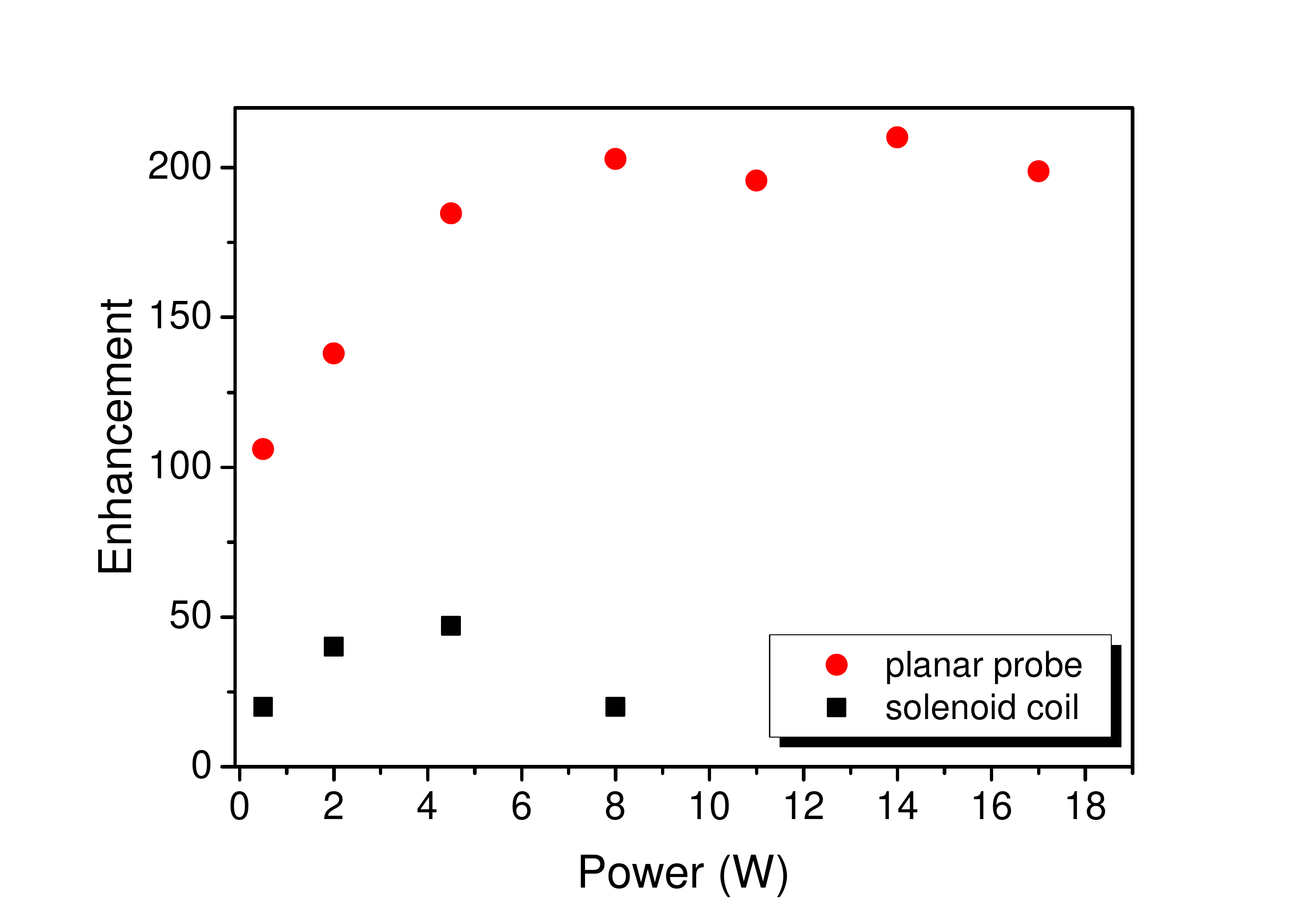}}
\caption{Overhauser ${}^{31}$P DNP enhancements of the liquid sample as a function of microwave power in the planar probe and in a solenoid coil.}\label{enhancement_power}
\end{figure}

\textbf{Planar probe.}~~As illustrated in Fig. \ref{Experiment-set-up} (b), the planar probe is composed of 1) a grid that consists of a thin layer of 150 parallel copper wires with a width of 50 $\mu$m and a period of 100 $\mu$m etched on a 300 $\mu$m-thick fused silica substrate, and 2) a 3 mm-thick copper support that has a square indentation (1 cm $\times$ 1 cm  $\times$ 100 $\mu$m) in which the sample is placed. The width and period of the wires of the grid were designed to have a high transmission of over 90 \% at about 260 GHz when the MW is polarized with the $\mathrm{E_1}$ field perpendicular to the wires. The rectangular and triangular strip contacts at both ends of the wires of the grid are designed so as to feed homogenous RF-currents to the wires. The rectangular and triangular strip contacts are connected to the support and to the central core of a coaxial cable, respectively. The grid and capacitors (not shown) form the RF resonator used for NMR. The grid sits on the support and is gently pressed down by three Cu-Be springs(Fig. \ref{Experiment-set-up} (b)). After clamping the grid and its support together, liquid samples can be injected from the holes at the bottom of the support into the indentation. The support is gold-coated to protect from chemical reactions with the sample.

 The support acts as a ground plane for MW and a heat sink. With nearly ideal boundary condition (i.e. $\mathrm{E_1}$ = 0) imposed by the conductive ground plane, the sample is located at a maximum of $\mathrm{B_1}$, which is a node of $\mathrm{E_1}$. This geometry results in reducing $\mathrm{E_1}$ and, hence, the dielectric heating in the sample. As shown in the simulation of $\mathrm{E_1}$ amplitude above the Cu support (COMSOL Multiphysics) in Fig. \ref{simulation} (a) and (b), $\mathrm{E_1}$ is minimized at the ground plane, and becomes stronger farther away from the Cu support. As the skin depth of pure water at 260 GHz is about 200 $\mu$m\cite{water_dielectric} larger than 100 $\mu$m of the sample thickness, all nuclear spins in the sample can be hyperpolarized by DNP. The sample is also in the thermal contact with the support that has a high thermal conductivity and a large thermal capacity. This enables rapid heat transport from the sample to the support. For the above two reasons, we expect to be able to apply high power MW without producing severe sample heating.
 
\textbf{Quasi optical units to produce circularly polarized microwaves.}~~The quasi optical $\lambda$/4 and $\lambda$/8 mirrors for 260 GHz are placed after the gyrotron output window, and are tailored to transform the linearly polarized gyrotron output to circular polarization. The circular polarization produced by the mirrors is twice as effective at saturating electron spins as linearly polarized MW. We previously showed that circularly polarized microwaves are able to induce a larger enhancement in frozen solutions than linearly polarized microwaves at the same power\cite{Yoon2016}. However, the grid  on the probe itself acts as a linear polarizer, rejecting half the power. An alternative probe configuration uses a coil loosely wound around the Cu support and the fused silica, in place of the wire grid. While this produces somewhat inhomogeneous $\mathrm{B_1}$ in the sample place, it is less polarization selective and the average $\mathrm{B_1}$ intensity is higher when circularly polarized MW beams are used. This method was used for ${}^1$H DNP, while the grid was used for ${}^{31}$P DNP.\\

\textbf{Liquid state ${}^1$H DNP at 9 T in water.}~~We obtained liquid state ${}^1$H DNP at 9.2 T (${}^1$H = 395 MHz) and at room temperature with water containing 80 mM TEMPOL (4-hydroxy-2,2,6,6-tetramethylpiperidin-1-oxyl). The DNP mechanism in liquid state is the Overhauser effect\cite{PhysRev.102.975}. Recently, the Overhauser effect was also observed in the solid phase of glass-forming media, such as ortho-tetraphenyl or polystrene, doped with radicals at high fields (over 9 T) and even at room temperature\cite{Can2014,Lelli2015}. Polarization transfer from electron to nuclear spins is mediated by cross relaxation due to time-dependent scalar or dipolar interactions. The coupling factor that represents the transfer efficiency becomes larger as the spectral density of the time-dependent interactions contains larger component near the EPR frequency. As the EPR frequency increases and becomes comparable to the inverse correlation time of translational or tumbling motions of the molecules, the component near the EPR frequency in the spectral density rapidly decreases. This implies that the coupling factor becomes smaller as the magnetic field increases, which results in less efficient liquid state DNP at high field\cite{Hofer2008,Prandolini2008}. In particular, the coupling factor between water protons and radicals via dipolar relaxation was predicted to be negligible at high field. However, Prisner \textit{et al.} observed surprisingly larger dipolar relaxation in water with ${}^{14}$N-TEMPOL at 9 T. They obtained a ${}^1$H enhancement of about -14 at about 40 ${}^{\circ}$C. The enhancement increased up to about -40 with temperature increasing up to 95 ${}^{\circ}$C as a result of faster diffusion and shorter correlation time of water molecules at higher temperatures\cite{Neugebauer2013}. Here, we irradiated circularly polarized MW of about 70 W, whose magnetic field strength is estimated to be about $\simeq$ 1.3 G in free space and $\simeq$ 2.5 G on the ground plane of the Cu support. In a previous report, the Prisner group obtained a saturation factor of over 0.9 in ${}^{14}$N-TEMPOL dissolved in water with a $\mathrm{B_1}$ of about 1.4 G (100 mW with a conversion factor of 0.45 mT$\cdot$W${}^{-1/2}$)\cite{Neugebauer2013}. Fig. \ref{simulation} (c) and (d) present our simulation results on the $\mathrm{B_1}$ distribution in the sample region of the planar probe with tan $\mathrm{\delta}$ = 0.8 (water at $\simeq$ 5 ${}^{\circ}$C) and tan $\mathrm{\delta}$ = 1.2  (water $\simeq$ 30 ${}^{\circ}$C), respectively\cite{water_dielectric}. The loss tangent increases rapidly from $\simeq$ 0.7 to $\simeq$ 1.4 with temperature in the range from 0 ${}^{\circ}$C to 50 ${}^{\circ}$C\cite{water_dielectric,Ellison2007}. Therefore, the sample heating by MW induces an increase in loss tangent, which results in shorter skin depth, and smaller $\mathrm{B_1}$. As shown in Fig. \ref{simulation} (c) for tan $\mathrm{\delta}$ = 0.8 with linearly polarized MW of 70 W, the $\mathrm{B_1}$ at the center shows a maximum of about 2.8 G right at the ground plane, and decreases down to 1.8 G at the bottom of the fused silica substrate. This implies that the $\mathrm{B_1}$ at the center in the middle of the probe is sufficiently large at any depth of the sample. $\mathrm{B_1}$ decreases laterally because of the Gaussian power distribution in the MW beam. Thus, $\mathrm{B_1}$ is insufficient to saturate the electron spins far away from the middle of the probe. Therefore, we used only 2 $\mu$L in order to place all the sample around the center, thus ensuring hyperpolarization of all the nuclei. This sample volume is about 700-fold greater than previously reported ($\sim$ 3 nL) using a microwave resonator for 260 GHz\cite{Neugebauer2013}. If the temperature (tan $\mathrm{\delta}$) increases to 30 ${}^{\circ}$C (1.2), the skin depth is expected to decrease down to 85 $\mu$m that is smaller than the sample thickness, most of the power is absorbed above the ground plane. This short skin depth causes the $\mathrm{B_1}$ maximum to occur underneath the grid, and also to reduce its strength to 1.9 G, as shown in Fig. \ref{simulation} (d). 
\begin{figure*}
\centerline{\includegraphics[width=17cm,height = 7cm]{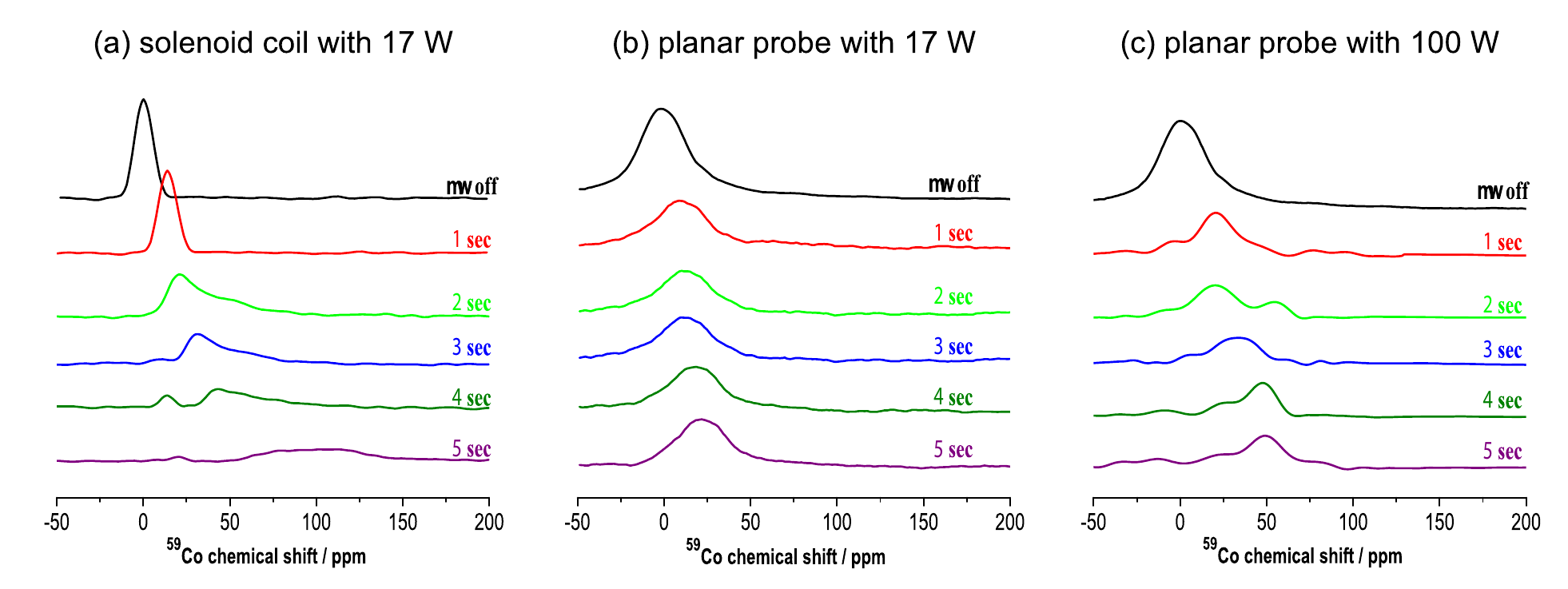}}
\caption{${}^{59}$Co NMR of $\mathrm{K_3[Co(CN_6)])}$ in water after exposing the room temperature liquid to millimeter power.}\label{Co_spectrum}
\end{figure*}

On the contrary, $\mathrm{E_1}$ is constantly minimized at the ground plane in both tan $\mathrm{\delta}$, as shown in Fig. \ref{simulation} (a) and (b). The $\mathrm{B_1}$ for tan $\mathrm{\delta}$ = 1.2 in most of the sample region is lower than that for tan $\mathrm{\delta}$ = 0.8, which implies that a larger power is required at higher temperature in order to saturate electron spins. Therefore, a loose NMR coil and circularly polarized MW were used for ${}^1$H DNP in this experiment in order to obtain saturation. The DNP frequency was set to the center of the electron paramagnetic resonance (EPR) spectrum of ${}^{14}$N-TEMPOL, which was obtained using our own high field EPR spectrometer\cite{Caspers2016}.

We turned on MW for 1 sec and turned off MW for 5 sec to cool the sample. Figure \ref{DNPspectrum} (a) shows the ${}^1$H Overhauser DNP spectrum and the NMR spectrum without MW. The negative enhancement indicates dipolar relaxation between electrons and ${}^1$H, consistent with previous results. The enhancement $\epsilon$ was calculated as $\epsilon = I/I_0 -1$, where $I_0$ and \textit{I} are the integrated areas of the DNP and NMR signals. The enhancement is estimated to be about -14, which is similar to the previous results that Prisner \textit{et al.} obtained at 40 ${}^{\circ}$C\cite{Neugebauer2013}. The large broadening about 10 ppm in the spectra comes from the inhomogeneity of the NMR magnet of $\sim$ 2 ppm/$\mathrm{mm}^2$. We were unable to use ${}^1$H NMR thermometry in order to estimate the sample temperature after MW irradiation because the ${}^1$H NMR frequency shift as a function of temperature (-0.012 ppm/${}^{\circ}$C)\cite{Gafurov2012} is much smaller than our NMR magnet inhomogeneity. 

However, the enhancement found, similar to that at 40 ${}^{\circ}$C in ref.\cite{Neugebauer2013}, implies a similar sample temperature after MW irradiation because we can assume that the electron spins were almost fully saturated in our case also. Our estimate of the temperature increase is also supported by the NMR thermometry experiments described below. The DNP signal intensity remains constant between experimental cycles, which also indicates that no sample loss occurs due to boiling.\\

\textbf{Liquid state ${}^{31}$P DNP at 9 T in fluorobenzene solution.}~~We also performed liquid state ${}^{31}$P DNP with a polar solution of triphenylphosphine ($\mathrm{Ph_3P}$) dissolved in fluorobenzene ($\mathrm{C_6H_5F}$) with 80 mM BDPA ($\alpha$,$\gamma$- bisdiphenylene-$\beta$-phenylallyl). The $\epsilon^{''}$ for $\mathrm{C_6H_5F}$ at 3 GHz was estimated to be 1.5 in a previous report\cite{Poley1955}, while that for water is about 29 at the same frequency. BDPA was reported to have a longer $T_{1e}$ than TEMPOL, so a smaller microwave power is needed. Griffin \textit{et al.} performed liquid state ${}^{31}$P DNP at 5 T with $\mathrm{Ph_3P}$ dissolved in a non-polar solvent of benzene with BDPA, and obtained a high enhancement of about 180 using 0.5 W of MW power\cite{Loening2002}. The solvent used here ($\mathrm{C_6H_5F}$) is polar, and much larger dielectric heating is expected than in the benzene solutions that Griffin \textit{et al.} used\cite{Loening2002}.  Figure \ref{DNPspectrum} (b) shows the DNP and NMR spectra obtained by the planar probe with 17 W and a sample volume of 10 $\mu$L. MW was irradiated at the center of the EPR spectrum of BDPA. A positive enhancement was observed, which indicates scalar relaxation consistent with ref\cite{Loening2002}. Although smaller enhancement was expected since we are in a higher field (9 T vs. 5 T), we obtained a similar enhancement of a factor of about 200, which shows effective scalar relaxation even at 9.3 T. Liquid state ${}^{31}$P DNP was also demonstrated in a standard solenoid coil with 2 mm inner diameter. While the planar probe allows high power MW irradiation without severe sample heating, the DNP signal with the solenoid coil had to be obtained with MW powers of less than 8 W in order to prevent sample loss due to boiling, as depicted in Fig. \ref{enhancement_power}. The enhancement in the solenoid coil increases up to about 50 at 4.5 W and begins to decrease with further power due to large sample loss. On the contrary, the enhancement obtained by the planar probe initially increases rapidly as the power increases, and stays constant around 200 beyond the saturation power of 8 W. The sample in the solenoid coil has smaller enhancement than that in the planar probe because only the fraction of the sample located within the skin depth can be directly enhanced. A solenoid coil with smaller diameter showed smaller enhancement than the solenoid coil used in this study because it had smaller heat capacity, causing even more severe sample loss.\\

\textbf{Measurement of sample temperature by NMR thermometer.}~~To quantify the decrease in dielectric heating in the planar probe, we performed NMR thermometry using potassium hexacyanocobaltate ($\mathrm{K_3[Co(CN_6)]}$) dissolved in water, as the ${}^{59}$Co chemical shift has been reported as having a large dependence on temperature($\approx$1.504 ppm/${}^{\circ}$C)\cite{Levy1980}. We obtained ${}^{59}$Co NMR spectra in the solenoid coil and the planar probe with increasing MW irradiation time from 1 sec to 5 sec. Figure \ref{Co_spectrum} (a) shows the ${}^{59}$Co NMR spectra obtained in the solenoid coil as a function of irradiation time with 17 W of MW power. The spectrum without MW in the solenoid coil has a broadening of about 10 ppm due to the magnetic field inhomogeneity of the NMR magnet used in this study, while that in the planar probe has a larger inhomogeneous broadening of about 25 ppm due to the larger sample area. The sample volume in the solenoid coil was about 40 $\mu$L. The spectrum is shifted by about 15 ppm at an irradiation time of 1 sec. As the irradiation time increases, the spectrum not only shows a larger shift, but also becomes broadened and distorted due to temperature gradients in the sample. The spectrum at an irradiation time of 5 sec is shifted to about 110 ppm, and becomes remarkably broadened, which indicates a severe sample heating and an inhomogeneous temperature in the sample. We found that more than half of the sample had evaporated after the experiment. On the contrary, the planar probe shows much smaller shifts and negligible broadening increase with irradiation time, as depicted in Fig. \ref{Co_spectrum} (b). A maximum shift of about 23 ppm is observed after an irradiation time of 5 sec without a significant broadening or distortion in the spectrum. This implies that dielectric heating is kept under control in the planar probe and that the temperature gradient remains negligible. 
\begin{figure}`
\centerline{\includegraphics[width=0.53\textwidth]{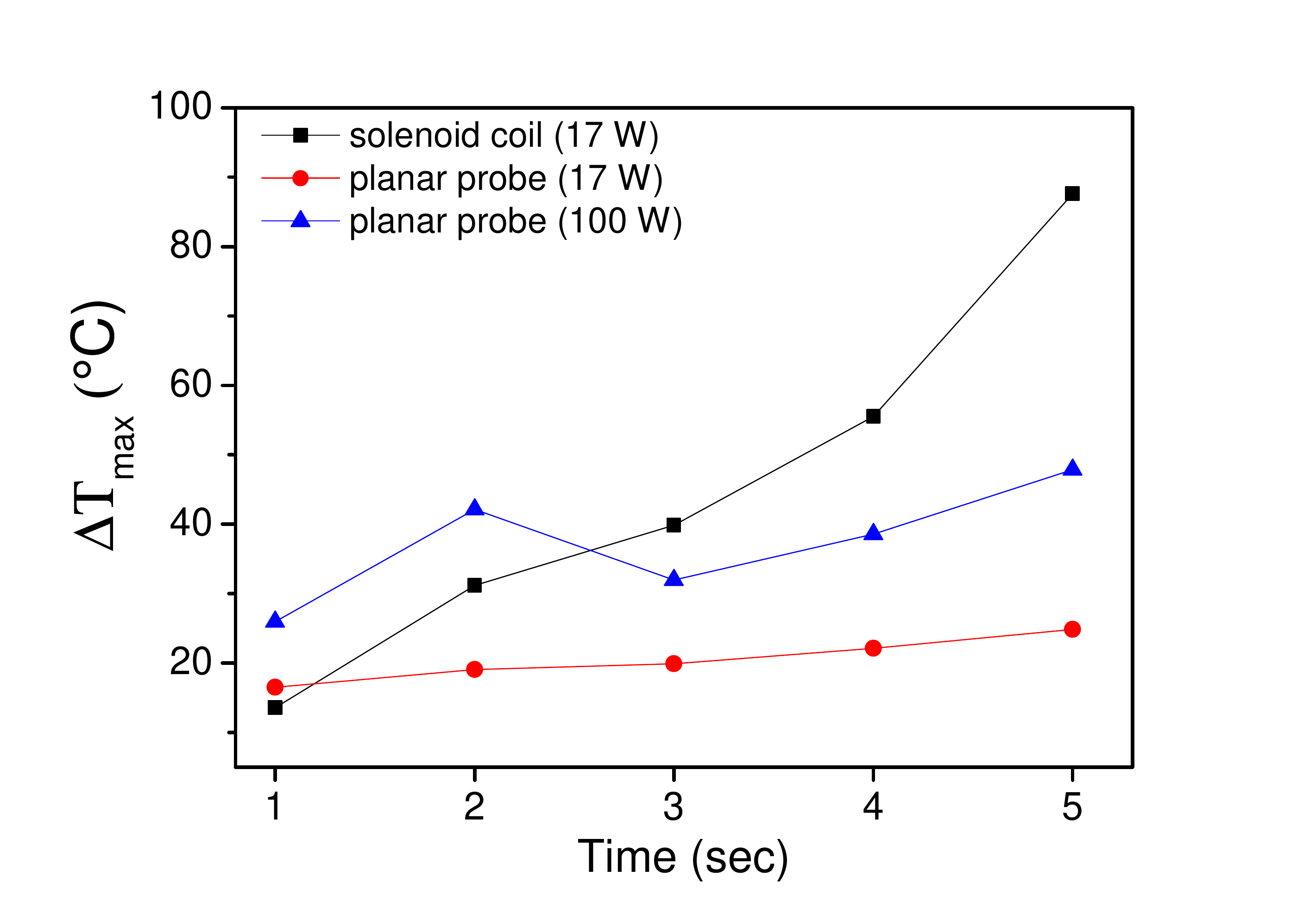}}
\caption{The maximum temperature increases in the planar probe and in the solenoid coil.}\label{maximum_temperature}
\end{figure}

We also tried a much higher MW power of 100 W. This MW power is strong enough to bring a frozen solution in a normal solenoid coil at 20 K to the liquid state at room temperature within 1.5 sec\cite{Yoon2016142}. Starting from a sample at room temperature, a large sample loss would occur in the solenoid coil due to evaporation even with a small irradiation time, so this experiment was performed only with the planar probe, as shown in  Fig. \ref{Co_spectrum} (c). The spectra show larger shifts compared to those in Fig. \ref{Co_spectrum} (b), and become more distorted because the temperature in the sample becomes inhomogeneous. However, the spectrum is shifted only to about 45 ppm ($\simeq$ 30${}^{\circ}$C) even with an irradiation time of 5 sec, showing an efficient suppression of dielectric heating. No sample loss was observed even with an irradiation time of 5 sec.     

Since a large temperature gradient occurs in the solenoid coil, we compared only maximum temperature increases in the planar probe and in the solenoid coil, as displayed in Fig. \ref{maximum_temperature}. The maximum temperature increases $\mathrm{\triangle T_{max}}$ are estimated by the positions at half maximum of the peak height. For the solenoid coil, some part of the sample were estimated to reach temperatures as high as 100 ${}^{\circ}$C, which is consistent with the large sample loss observed after the experiment. On the other hand, the temperature of the sample in the planar probe is found to increase by about 20 ${}^{\circ}$C with 17 W of MW power. Even at 100 W, the planar probe shows an increase of only about 25 ${}^{\circ}$C for an irradiation time of 1 sec. We expect a roughly similar increase in the sample temperature after MW irradiation of 70 W in the liquid state ${}^1$H DNP.

\section{Discussion}
In this study, we showed a new methodology for increasing the sample volume for liquid state DNP at 9 T from nL to $\mu$L by using a high power gyrotron and a planar probe. The increase in sample volume has several benefits; first, easy sample handling (injection into the planar probe, or dropping on the Cu support), second, negligible interference from background signals, and third, ${}^1$H NMR signal intensity large enough for a single scan. Even though sufficient MW power was produced by the gyrotron, the ${}^1$H enhancement was limited by the small coupling factor. The enhancement could be improved by using radicals with narrower EPR lines such as ${}^{15}$N-TEMPOL and Fremy's Salt\cite{Prandolini2009,Gafurov2012} or using supercritical fluids, whose correlation time for molecular motion would be much shorter than that of water\cite{VanBentum2016,Wang2015}. Dielectric heating in the planar probe can be further reduced by using thinner samples while still holding $\mu$L of liquid. This new method can be extended to higher or lower magnetic fields by modifying the thickness of the sample region and the grid. 

\section{Methods}
\textbf{Sample preparation} The sample for liquid state ${}^1$H DNP was prepared by dissolving 80 mM ${}^{14}$N-TEMPOL (Sigma Aldrich) in de-ionized water without degassing. The prepared sample of about 2 $\mu$L was dropped onto the center of the Cu support and closed by a silica cover. The sample for liquid state ${}^{31}$P DNP was prepared in the following manner: first, $\mathrm{N_2}$ was bubbled in fluorobenzene in a glove box, second, a 2 M $\mathrm{Ph_3P}$ solution was dissolved  in the $\mathrm{N_2}$ bubbled-fluorobenzene, and third, 80 mM BDPA (Sigma Aldrich) was dissolved in the solution prepared at the preceded step. 1M $\mathrm{K_3[Co(CN_6)]}$ dissolved in de-ionized water was used for the ${}^{59}$Co NMR thermometry experiments.\\

\begin{acknowledgments}
We gratefully acknowledge financial support by the Swiss National Science Foundation, Requip (No.
206021-121303/1), FN($\mathrm{200021{-}153230}$), and CTI-Project no. 15617.1 PFNM-NM.
\end{acknowledgments}

\section{Author contributions}
D.Y. carried out the DNP experiments, A.D. and E.D.R. contributed with simulations, C.C. and M.S. with high field EPR measurements, J.G. and S. A. implemented the gyrotron settings, D.Y. and M.S. prepared the samples, D.Y. and J-Ph. A. designed the experiment.


\begin{thebibliography}{50}

\bibitem{Abragam}A. Abragam, Principles of Nuclear Magnetism, INTERNATIONAL SERIES OF MONOGRAPHS ON PHYSICS:32 (OXFORD UNIVERSITY PRESS, 1960).

\bibitem{Armstrong2009}B. D. Armstrong and S. Han, J. Am. Chem. Soc. 131, 4641 (2009).

\bibitem{Hofer2008}P. H\"{o}fer, G. Parigi, C. Luchinat, P. Carl, G. Guthausen, M. Reese, T. Carlomagno, C. Griesinger, and M. Bennati, J. Am. Chem. Soc. 130, 3254 (2008).

\bibitem{Tuerke2012}M.-T. Tuerke, G. Parigi, C. Luchinat, and M. Bennati, Phys. Chem. Chem. Phys. 14, 502 (2012).

\bibitem{Becerra1993}L. Becerra, G. Gerfen, R. Temkin, D. Singel, and R. Griffin, Phys. Rev. Lett. 71, 3561 (1993).

\bibitem{Rosay2010}M. Rosay, L. Tometich, S. Pawsey, R. Bader, R. Schauwecker, M. Blank, P. M. Borchard, S. R. Cauman, K. L. Felch, R. T. Weber, R. J. Temkin, R. G. Griffin, and W. E. Maas, Phys. Chem. Chem. Phys. 12, 5850 (2010).

\bibitem{Kemp2016}T. F. Kemp, H. R. W. Dannatt, N. S. Barrow, A. Watts, S. P. Brown, M. E. Newton, and R. Dupree, J. Magn. Reson. 265, 77 (2016).

\bibitem{Bruker}www.bruker.com/products/mr/nmr/dnp-nmr/overview.html.

\bibitem{Thurber2010}K. R. Thurber, W.-m. Yau, and R. Tycko, J. Magn. Reson. 204, 303 (2010).

\bibitem{Nanni2011}E. a. Nanni, A. B. Barnes, Y. Matsuki, P. P. Woskov, B. Corzilius, R. G. Griffin, and R. J. Temkin, J. Magn. Reson. 210, 16 (2011).

\bibitem{Hu2004}K.-N. Hu, H.-h. Yu, T. M. Swager, and R. G. Griffin, J. Am. Chem. Soc. 126, 10844 (2004).

\bibitem{Zagdoun2013}A. Zagdoun, G. Casano, O. Ouari, M. Schw\"{a}rzwalder, A. J. Rossini, F. Aussenac, M. Yulikov, G. Jeschke, C. Coperet, A. Lesage, P. Tordo, and L. Emsley, J. Am. Chem. Soc. 135, 12790 (2013).

\bibitem{Prandolini2009}M. J. Prandolini, V. P. Denysenkov, M. Gafurov, B. Endeward, and T. F. Prisner, J. Am. Chem. Soc. 131, 6090 (2009).


\bibitem{Prandolini2008}M. J. Prandolini, V. P. Denysenkov, M. Gafurov, S. Lyubenova, B. Endeward, M. Bennati, and T. F. Prisner, Appl. Magn. Reson 34, 399 (2008).

\bibitem{Gafurov2012}M. Gafurov, V. Denysenkov, M. J. Prandolini, and T. F. Prisner, Appl. Magn. Reson. 43, 119 (2012).

\bibitem{Neugebauer2013}P. Neugebauer, J. G. Krummenacker, V. P. Denysenkov, G. Parigi, C. Luchinat, and T. F. Prisner, Phys. Chem. Chem. Phys. 15, 6049 (2013).

\bibitem{annino2013}G. Annino, A. Macor, Emile De Rijk, Stefano Alberti, Magnetic resonance hyperpolarization and multiple irradiation probe head," (2013), wO Patent App. PCT/EP2012/062492.

\bibitem{Alberti2012}S. Alberti, J. P. Ansermet, K. a. Avramides, F. Braunmueller, P. Cuanillon, J. Dubray, D. Fasel, J. P. Hogge, a. MacOr, E. De Rijk, M. Da Silva, M. Q. Tran, T. M. Tran, and Q. Vuillemin, Phys. Plasmas 19 (2012).

\bibitem{Alberti2013}S. Alberti, F. Braunmueller, T. M. Tran, J. Genoud, J. P. Hogge, M. Q. Tran, and J. P. Ansermet, Phys. Rev. Lett. 111, 1 (2013).

\bibitem{Yoon2016}D. Yoon, M. Soundararajan, P. Cuanillon, F. Braunmueller, S. Alberti, and J. P. Ansermet, J. Magn. Reson. 262, 62 (2016).

\bibitem{water_dielectric}S. K. Misra, Multifrequency epr: Experimental considerations, Wiley-VCH Verlag GmbH \& Co. KGaA, (2011) pp. 229-294.

\bibitem{PhysRev.102.975}T. R. Carver and C. P. Slichter, Phys. Rev. 102, 975 (1956).

\bibitem{Can2014}T. V. Can, M. A. Caporini, F. Mentink-Vigier, B. Corzilius, J. J. Walish, M. Rosay, W. E. Maas, M. Baldus, S. Vega, T. M. Swager, and R. G. Griffin, J. Chem. Phys. 141, 064202 (2014).

\bibitem{Lelli2015}M. Lelli, S. R. Chaudhari, D. Gajan, G. Casano, A. J. Rossini, O. Ouari, P. Tordo, A. Lesage, and L. Emsley, J. Am. Chem. Soc. 137, 14558 (2015).

\bibitem{Ellison2007}W. J. Ellison, Journal of Physical and Chemical Reference Data 36, 1 (2007).

\bibitem{Caspers2016}C. Caspers, P. F. da Silva, M. Soundararajan, M. A. Haider, and J.-P. Ansermet, APL Photonics 1, 026101 (2016).

\bibitem{Poley1955}J. P. Poley, Appl. Sci. Res. 4, 337 (1955).

\bibitem{Loening2002}N. M. Loening, M. Rosay, V. Weis, and R. G. Griffin, J. Am. Chem. Soc. 124, 8808 (2002).

\bibitem{Levy1980}G. C. Levy, J. Terry Bailey, and D. A. Wright, J. Magn. Reson. 37, 353 (1980).

\bibitem{Yoon2016142}D. Yoon, M. Soundararajan, C. Caspers, F. Braunmueller, J. Genoud, S. Alberti, and J.-P. Ansermet, J. Magn. Reson. 270, 142 (2016).

\bibitem{VanBentum2016}J. van Bentum, B. van Meerten, M. Sharma, and A. Kentgens, J. Magn. Reson. 264, 59 (2016).


\bibitem{Wang2015}X. Wang, W. C. I. Iii, S. I. Salido, Z. Sun, L. Song, K. H. Tsai, and C. J. Cramer, Chem. Sci. 6, 6482 (2015).
	
\end{thebibliography}
\end{document}